%% file: Main_Paper_ACE_Challenge_2015.tex
\documentclass{article}
\usepackage{ace,url,times}
\usepackage{amssymb,amsmath,amsfonts}
\usepackage{graphicx}  
\usepackage{psfrag}     
\usepackage{xspace} 
\usepackage{acronym}
\usepackage{color}

\newcommand{\draftplot}{false}
\newcommand{\fwidth}{1\columnwidth}  
\newcommand{\vspalab}{\vspace{-4ex}} 

\newcommand{\presec}{\vspace*{-0.2mm}}  
\newcommand{\postsec}{\vspace*{-2.9mm}} 

\definecolor{ccyan}{rgb}{0.49,1.0,1.0}
\definecolor{cpurple}{rgb}{0.49,0.49,1.0}
\definecolor{cgreen}{rgb}{0.49,1.0,0.49}
\definecolor{cred}{rgb}{0.88,0.49,0.49}

\input{macros_ACE2015}
\input{acronyms_ACE2015}

\title{Single-Channel Maximum-Likelihood T60 Estimation\\ Exploiting Subband Information}

\twoauthors
  {Heinrich W. L\"ollmann, Andreas Brendel, Walter Kellermann}
    {\normalsize Friedrich-Alexander University Erlangen-N\"urnberg (FAU)\\
\normalsize Multimedia Communications and Signal Processing (LMS)\\
\normalsize 91058 Erlangen, Germany \\
 \texttt{\normalsize heinrich.loellmann@fau.de, andreas.brendel@fau.de,} \\
 \texttt{\normalsize walter.kellermann@fau.de}}
  {Peter Vary}
    {\normalsize RWTH Aachen University  \\
    \normalsize Institute of Communication Systems (IND)\\
    \normalsize 52056 Aachen, Germany\\
  \texttt{\normalsize peter.vary@ind.rwth-aachen.de  }\\
}

\begin{document}

\ninept
\maketitle

\begin{sloppy}

\input{content_paper_ACE_Challenge_2015}

\end{sloppy}


\bibliographystyle{IEEEtran} 
\bibliography{my_references_20150612}


\end{document}

%% file: macros_ACE2015.tex
\newcommand{\fott}[1]{}   
\newcommand{\sparaise}[3]{\hspace{#1}\raisebox{#2}{#3}} 

\newcommand{\eref}[1]{\mbox{Eq.\hspace{0.5ex}\eqref{#1}}} 
\newcommand{\Eref}[1]{\mbox{Eq.\hspace{0.5ex}\eqref{#1}}}
\newcommand{\fref}[1]{\mbox{Fig.\hspace{0.5ex}\ref{#1}}}
\newcommand{\frefs}[1]{\mbox{Figs.\hspace{0.5ex}\ref{#1}}}
\newcommand{\tabref}[1]{\mbox{Table\hspace{0.5ex}\ref{#1}}}
\newcommand{\Tabref}[1]{\mbox{Table\hspace{0.5ex}\ref{#1}}}
\newcommand{\secref}[1]{\mbox{Sec.\hspace{0.5ex}\ref{#1}}}
\newcommand{\labeling}[2]{\caption{\textbf{\label{#1}}\textit{#2}}} 
\newcommand{\cf}{cf.,\xspace}
\newcommand{\eg}{e.g.,\xspace}
\newcommand{\qmarks}[1]{`#1'}

\newcommand{\mequal}{\stackrel{!}{=}}                       
\newcommand{\mul}{\hspace{0.3ex}}                           
\newcommand{\myforall}{\hspace{0.5ex}\forall\hspace{0.5ex}}
\newcommand{\eqdot}{\hspace{1ex}.}                          
\newcommand{\eqspace}{\hspace{2ex}}                         
\newcommand{\eqtext}[1]{\eqspace\text{#1}\eqspace}
\newcommand{\expec}[1]{E\left\{#1\right\}}           
\newcommand{\maximum}[2]{\underset{#1}{\text{max}}\left\{#2\right\}}
\newcommand{\minimum}[2]{\underset{#1}{\text{min}}\left\{#2\right\}}
\newcommand{\mmax}[1]{\text{max}\left\{#1\right\}}
\newcommand{\mmin}[1]{\text{min}\left\{#1\right\}}
\newcommand{\setnat}{\mathbb{N}}  

\newcommand{\rirm}{d_\text{m}} 
\newcommand{\rirmt}{\tilde{d}_\text{m}} 
\newcommand{\rt}{T_{60}}  
\newcommand{\rte}{\widehat{T}_{60}} 
\newcommand{\rtedct}{\widehat{T}_{60}^\text{(dct)}} 
\newcommand{\rteoct}{\widehat{T}_{60}^\text{(oct)}} 
\newcommand{\rteml}{\widehat{T}_{60}^{\text{(ML)}}} 
\newcommand{\rtebase}{\widehat{T}_{60}^{\text{(base)}}} 
\newcommand{\rhoeml}{\hat{\rho}^{\text{(ML)}}} 
\newcommand{\ydec}{d}  
\newcommand{\rtfd}{\mathcal{T}_{60}(\mu)} 
\newcommand{\rtfdedct}{\widehat{\mathcal{T}}_{60}^\text{(dct)}(\mu)} 
\newcommand{\rtfdeoct}{\widehat{\mathcal{T}}_{60}^\text{(oct)}(\mu)} 
\newcommand{\rtfdemod}{\widehat{\mathcal{T}}_{60}^\text{(model)}(\mu)} 

\newcommand{\fs}{f_\text{s}}         
\newcommand{\Ts}{T_\text{s}}         
\newcommand{\ik}{k}           
\newcommand{\secs}{\,\text{s}}
\newcommand{\db}{\text{\,dB}\xspace}
\newcommand{\khz}{\text{\,kHz}\xspace}
\newcommand{\fb}{filter-bank\xspace}
\newcommand{\wrt}{w.r.t.\hspace{-0.2ex}\xspace}

%% file: acronyms_ACE2015.tex
\acrodef{ACE}{Acoustic Characterisation of Environments}
\acrodef{ASR}{automatic speech recognition}
\acrodef{DCT}{discrete cosine transform}
\acrodef{DFT}{discrete Fourier transform}
\acrodef{DRR}{direct-to-reverberant energy ratio}
\acrodef{RIR}{room impulse response}
\acrodef{RT}{reverberation time}
\acrodef{RTE}{reverberation time estimation}
\acrodef{ML}{maximum-likelihood}
\acrodef{MLE}{maximum-likelihood estimation}
\acrodef{MMSE}{minimum mean-square error}
\acrodef{SNR}{signal-to-noise-ratio}
\acrodef{PDF}{probability density function}
\acrodef{PSD}{power spectral density}

%% file: content_paper_ACE_Challenge_2015.tex
\begin{abstract}
This contribution presents four algorithms developed by the authors for single-channel fullband and subband T60 estimation within the ACE challenge. 
The blind estimation of the fullband reverberation time (RT) by maximum-likelihood (ML) estimation based on [15] is considered as baseline approach. An improvement of this algorithm is devised where an energy-weighted averaging of the upper subband RT estimates is performed using either a DCT or 1/3-octave filter-bank. The evaluation results show that this approach leads to a lower variance for the estimation error in comparison to the baseline approach at the price of an increased computational complexity. Moreover, a new algorithm to estimate the subband RT is devised, where the RT estimates for the lower octave subbands are extrapolated from the RT estimates of the upper subbands by means of a simple model for the frequency-dependency of the subband RT. The evaluation results of the ACE challenge reveal that this approach allows to estimate the subband RT with an estimation error which is in a similar range as for the presented fullband RT estimators.
\end{abstract}


\presec
\section{Introduction}
\label{sec:intro}
\postsec

The \ac{RT} $\rt$ is an important quantity to characterize the acoustical properties of an enclosure \cite{Kuttruff00}. Furthermore, knowledge about the \ac{RT} can be exploited for enhanced \ac{ASR}, \eg \cite{Yoshioka2012,Maas2012}, as well as speech dereverberation, \eg \cite{Lebart01,Habets07b,Loellmann09c,Loellmann09f}. For such applications, the \fott{possibly time-varying} \ac{RT} can usually not be determined from a known \ac{RIR} \cite{Schroeder65}, but has to be estimated \emph{blindly} from a reverberant speech signal, which is frequently also distorted by noise. Especially the above-mentioned applications have fueled the research interest in blind \ac{RTE} and numerous methods were proposed in recent years \cite{Ratnam03,Ratnam04,Wu06,Wen08,Falk08,Loellmann08c,Loellmann2010b,Prego2012,Lopez2012,Talmon2013,Diether2015,Schluedt2015}. The variety of concepts raises the desire for an \emph{objective} comparison of different algorithms for $\rt$ estimation as published in \cite{Gaubitch2012}. A more comprehensive comparison of algorithms for \ac{RTE} as well as \ac{DRR} estimation is facilitated by the \acfi{ACE}\acused{ACE} challenge \cite{Eaton2015}. The participants had access to a development (Dev) database and an evaluation (Eval) database. The single-channel Dev database comprises 288\fott{ different} noisy and reverberant speech files for which the ground-truth of the fullband and subband \ac{RT}, the \ac{SNR}, as well as the \ac{DRR} is provided to allow the participants to develop and tune their algorithms. The single-channel Eval database comprises 4500 speech files without ground-truth data for generating the submission results for the challenge.

This contribution provides a description of the four algorithms developed by the authors for the submission of their results for single-channel fullband and subband $\rt$ estimation. 

All the presented algorithms for \ac{RTE} employ a \acfi{ML}\acused{ML} estimation. The use of an \ac{ML} estimator for blind single-channel \ac{RTE} was first presented in \cite{Ratnam03,Ratnam04}. In \cite{Loellmann08c}, this concept is extended to estimate the \ac{RT} from a noisy \ac{RIR} and to estimate the \ac{RT} blindly from a noisy and reverberant speech signal. A further development of this algorithm is presented in \cite{Loellmann2010b} to allow for a fast tracking of time-varying \acp{RT} with low complexity from noisy and reverberant speech signals. A slightly modified version of this algorithm for single-channel fullband $\rt$ estimation has been employed as baseline algorithm for the \ac{ACE} challenge and is described in \secref{sec:baseline}. An improvement of this algorithm by averaging subband \ac{RT} estimates has been developed\fott{ for the \ac{ACE} challenge}, which is described in \secref{sec:subband_averaging}. Moreover, results for the blind estimation of the subband \ac{RT} have 
been submitted, and the algorithm devised for this is presented in \secref{sec:subband_RTE}. The evaluation results of the \ac{ACE} challenge are discussed in \secref{sec:results} and the paper concludes with \secref{sec:conclusions}.

\presec
\section{Baseline Algorithm}
\label{sec:baseline}
\postsec

The baseline algorithm employed for fullband \ac{RTE} is a slightly modified version of the algorithm presented in \cite{Loellmann2010b}. It is referred to as \emph{baseline algorithm} since the Matlab code has been published on Matlab Central \cite{jeub_2012_matlab_exhange}. In contrast to \cite{Loellmann2010b}, the fast tracking of time-varying \acp{RT} is omitted here to obtain more robust estimates. 

\subsection{Model for ML Estimation}
It is assumed that the reverberant speech signal is obtained by a speech signal $s(\ik)$ convolved with a time-varying \ac{RIR} $h(\eta,\ik)$ of \fott{(possibly
infinite)} length $L_h$:
\begin{align}
  z(\ik) = \sum\limits_{\eta=0}^{L_h-1} s(\ik-\eta) \cdot h(\eta,\ik) ,
\end{align}
with $k$ denoting the discrete time index. If a speech pause begins
\begin{align}
  s(k-\eta) \begin{cases}
     \approx0 &\eqtext{for} \eta=0,1,\ldots,L_\text{o}-1
     \\ \neq 0 & \eqtext{for} \eta=L_\text{o},\ldots,L_h-1,
  \end{cases}
\end{align}
and the room reverberation causes a decaying signal $\ydec(\ik)$ since
\begin{align}
 \label{eq:def_sound_decay}
  z(\ik) =
     \underbrace{\sum\limits_{\eta=0}^{L_\text{o}-1}
          s(\ik-\eta) \cdot h(\eta,\ik)
          }_{\displaystyle \approx 0}
          +
     \underbrace{\sum\limits_{\eta=L_\text{o}}^{L_h-1}
       s(\ik-\eta) \cdot h(\eta,\ik)
       }_{ \displaystyle = \ydec(\ik)} 
,
\end{align}
assuming that $h(\eta,k)\neq0$ for at least one value $L_\text{o}\leq \eta < L_h$. The sound decay $\ydec(\ik)$ is \emph{modeled} by a discrete random process
\begin{equation}
 \label{eq:RIR_model}
  \rirm(\ik) =  A_\text{r} \mul v(\ik) \mul e^{-\rho \mul k \mul T_\text{s} } \mul
  \epsilon(\ik)
\end{equation}
with real amplitude $A_\text{r}>0$, decay rate $\rho$ and unit step sequence $\epsilon(\ik)$. 
$\Ts=1/\fs$ denotes the sampling period and $v(\ik)$ is a sequence of i.i.d. random variables with normal distribution $\mathcal{N}(0,1)$. \Eref{eq:RIR_model} can also be seen as a simple statistical model for the \ac{RIR}, which considers only the effects of late reflections and models it as diffuse noise. The energy decay curve for the corresponding continuous-time sound decay model is given by the expectation
\begin{align}
   E_{\tilde{d}}(t) = \expec{ \rirmt^2(t) }
        = A_\text{r}^2 \mul  e^{- 2\mul \rho \mul t }
          \mul \tilde{\epsilon}(t)
\end{align}
where the tilde indicates the continuous-time counterparts to the discrete-time quantities of \eref{eq:RIR_model}.  A relation between \emph{decay rate} $\rho$ and \emph{reverberation time} $\rt$ can be established by the requirement
\begin{align}
  10\mul \log_{10} \left( \frac{ E_{\tilde{d}}(0) }{ E_{\tilde{d}}(\rt) } \right) \mequal 60
\end{align}
such that
\begin{align}
 \label{eq:rel_rt_decay_rate}
 \rt  = \frac{ 3 }{\rho\mul \log_{10}( e ) }
             \approx \frac{ 6.908}{ \rho } \eqdot
\end{align}
Due to this relation, the terms decay rate and \ac{RT} will be used interchangeably in the following.

According to the model given by \eref{eq:RIR_model}, the signal value $\ydec(\ik)$ of \eref{eq:def_sound_decay} is represented by a random variable with Gaussian \ac{PDF}
\begin{align}
  p_d(x,\ik) & = \frac{1}{\sqrt{ 2\mul\pi }\mul  \xi(\ik)}
   \mul \exp\left\{ -\frac{ x^2 }{ 2 \mul \xi^2(\ik) } \right\} 
\\ 
  \xi(\ik) & = A_\text{r} \mul a^{\ik} \mul \epsilon(k) 
   \eqtext{with}  a = e^{- \Ts \mul \rho } 
   .
\end{align}
The sequence $\ydec(\ik)$ for $\ik\in\{0,\ldots,N-1\}$ is then given by $N$ independent, normally distributed random variables with zero mean and non-identical \acp{PDF}. This allows to derive an \ac{ML} estimator for the unknown decay rate or \ac{RT}, respectively, \cite{Ratnam03,Loellmann08c}, where the decay rate $\rho$ is estimated from a given sound decay $\ydec(\ik)$ by finding the maximum
\begin{subequations} 
 \label{eq:ml_rte}
\begin{align}
  \rhoeml = \text{arg}\hspace{0.5ex} \maximum{\rho}{ \mathcal{L(\rho)}  }
\end{align}
of the log-likelihood function
\begin{align}
 \mathcal{L}(\rho) & =  \\ \notag
   - & \frac{N}{2} \left(
      (N-1) \ln( a )
      + \ln\left(
         \frac{2\mul\pi}{N} \sum\limits_{i=0}^{N-1} a^{-2\mul i} \ydec^2(i)
         \right)+1
          \right).
\end{align}
\end{subequations}
The corresponding estimate for the \ac{RT} $\rteml$ is given by \eref{eq:rel_rt_decay_rate}.

\subsection{Implementation}

In a first step, a speech denoising is performed by spectral weighting. For this, the overlap-add scheme for the \ac{DFT} is employed with half-overlapping frames  of 512 samples weighted with a von Hann window. The spectral weights are calculated by the spectral \ac{MMSE} estimator for the magnitude of the speech \ac{DFT} coefficients \cite{Erkelens07} where the noise \ac{PSD} is calculated by the \ac{MMSE}-based estimator presented in \cite{Hendriks2010}. A crucial parameter is the minimum value for the spectral weights $W_\text{min}$, the so-called noise floor, since it controls the reduction of noise and diffuse reverberation alike. For the baseline algorithm, a frequency-independent value of $W_\text{min}=0.2$ was found suitable.

The denoised signal $\hat{z}(\ik)$ is processed in signal frames of $M$ samples shifted by $M_\Delta$ sample instants
\begin{align}
  x_\text{f}(\lambda,m) = \hat{z}(\lambda\mul M_\Delta + m )
  \eqtext{with} m=0,1,\ldots,M-1
\end{align}
and frame index $\lambda\in\setnat$, where no downsampling is employed here in contrast to \cite{Loellmann2010b}.

In a first step, a pre-selection is conducted to detect potential frames with sound decays, \cf \eref{eq:def_sound_decay}. For this, the current frame $x_\text{f}(\lambda,m)$ is divided into $L=M/P\in\setnat$ sub-frames of length $P$
\begin{align}
  y(\lambda,l,\kappa) = x_\text{f}(\lambda,l \mul P + \kappa)
\end{align}
with $\kappa\in\{\,0,1,\ldots, P-1\,\}$ and $l\in\{\,0,1,\ldots, L-1\,\}$. It is then checked for frame $\lambda$ whether the energy, maximum and minimum value of a sub-frame $l$ deviates from the successive sub-frame $l+1$ according to
\begin{subequations}
  \label{eq:cond_decay_detection}
\begin{align}
 \sum\limits_{\kappa=0}^{P-1}  y^2(\lambda,l,\kappa)
  & >  w\cdot \sum\limits_{\kappa=0}^{P-1}  y^2(\lambda,l+1,\kappa)
  \\
  \mmax{ y(\lambda,l,\kappa) }  & > w \cdot \mmax{ y(\lambda,l+1,\kappa) }
 \\  
 \mmin{ y(\lambda,l,\kappa) }
   & <  w \cdot \mmin{ y(\lambda,l+1,\kappa) }
\end{align}
\end{subequations}
with sub-frame counter $l=0,1,\ldots, L-2$ and weighting factor $0 < w \leq 1$. If the conditions of \eref{eq:cond_decay_detection} are fulfilled for $l\geq l_\text{min}-1$ consecutive sub-frames, a possible sound decay is detected. In this case, the \ac{RT} is calculated from the signal segment of these $l$ consecutive sub-frames by means of \eref{eq:ml_rte} for a finite set of \ac{RT} values (decay rates), which is here given by $\rteml/\secs\in\{\,0.2,\,0.3,\ldots,1.1\,\}$. Otherwise, no sound decay is detected and the next signal frame $ y(\lambda+1,l,\kappa)$ is processed.

A new \ac{ML} estimate is used to update a histogram calculated for the last $K_\text{f}$ \ac{ML} estimates of the \ac{RT}. The \ac{RT} value associated with the maximum of the histogram is taken as preliminary estimate $\rte^{(1)}(\lambda)$.
The variance for the estimated \ac{RT} is reduced by recursive smoothing
\begin{align}
  \label{eq:rt_smoothed}
  \rte(\lambda) = \beta \cdot \rte(\lambda-1)
    + (1-\beta) \cdot\rte^{(1)}(\lambda)
\end{align}
with $0<\beta<1$. For the \ac{ACE} challenge, time-invariant \acp{RT} are considered such that the final estimate is obtained by the average
\begin{align}
  \rtebase= \frac{1}{n_2-n_1+1}\sum\limits_{\lambda=n_1}^{n_2} \rte(\lambda) 
\end{align}
with $n_1$ and $n_2$ denoting the first and last frame for which an \ac{RT} has been calculated. \Tabref{table:parameters_baseline_algorithm} lists the algorithmic parameters of the described baseline algorithm used for the \ac{ACE} challenge submission.
\begin{table}[tbh]
\centering
\begin{tabular}{c|c|c|c|c|c|c|c|c}
                     $\fs$  & $M$  & $M_\Delta$ & $P$ & $l_\text{min}$ & $L$ & $w$ & $K_\text{f}$ & $\beta$  \\[0.3ex] \hline
 \vphantom{$2^{2^2}$}16\khz & 4923 & 137        & 547 &       3        &  9  &   1   &   800        & 0.95  
\end{tabular}
\vspace{-1.5ex}
 \labeling{table:parameters_baseline_algorithm}{Algorithmic parameters of the baseline algorithm.}
\end{table}

\presec
\section{Subband RT Averaging}
\label{sec:subband_averaging}
\postsec

The baseline algorithm has been extended to improve the estimation accuracy by averaging subband \ac{RT} estimates. The concept of averaging subband \ac{RT} estimates has already been proposed in \cite{Prego2012}, but a very different realization of this concept is suggested here. 

The denoised signal $\hat{z}(\ik)$ is split into $N_\text{dct}$ subband signals $x_\mu(\ik)$ with subband index $\mu\in\{\,1,\ldots, N_\text{dct}\,\}$ by a uniform \ac{DCT}-IV \fb without downsampling as described in \cite{Loellmann11a}, where the FIR prototype filter of length $L_\text{dct}$ is designed according to \cite{Lin98} for a stopband attenuation of $100\db$. 

The estimate for the subband \ac{RT} $\rtfdedct$ is calculated by the previously described baseline algorithm for the upper subbands $n_\text{dct},\ldots,N_\text{dct}$. The fullband estimate is obtained by the weighted average
\begin{align}
 \rtedct= \frac{1}{N_\text{dct}-n_\text{dct}+1} \sum\limits_{\mu=n_\text{dct}}^{N_\text{dct}} w_\mu \mul \rtfdedct \eqdot
\end{align}
The weighting factors are given by the ratio of the subband signal energy to the sum of all considered subband energies 
\begin{align}
 w_\mu & = \frac{1}{E_0}\sum\limits_{\ik} x_\mu^2(\ik) \eqtext{for} \mu \in \{n_\text{dct},\ldots,N_\text{dct}\}
\\
E_0  & =\sum\limits_{\mu=n_\text{dct}}^{N_\text{dct}} \sum\limits_{\ik} x_\mu^2(\ik)  \eqdot
\end{align}
This weighting is motivated by the rationale that subband signals with a higher energy provide on average more reliable \ac{RT} estimates than subband signals with a low energy. This concept was confirmed by experiments with the Dev database where the weighting has led to superior results in comparison to a non-weighted averaging\fott{ ($w_\mu\equiv 1$)} (as well as the baseline approach). \tabref{table:parameters_DCT_algorithm} lists the algorithmic parameters used for the submission to the \ac{ACE} challenge.
\begin{table}[t!]
\centering
\begin{tabular}{c|c|c|c}
$N_\text{dct}$           & $n_\text{dct}$ & $L_\text{dct}$  & $W_\text{min}$ \\[0.3ex] \hline
 \vphantom{$2^{2^2}$} 12 & 3              & 174             &  0.35  
\end{tabular}
\vspace{-1.5ex}
 \labeling{table:parameters_DCT_algorithm}{Algorithmic parameters for fullband \ac{RTE} with subband RT averaging by means of a \ac{DCT}-IV \fb.}
\end{table}

The described concept for fullband \ac{RTE} by subband RT averaging has also been implemented with a 1/3-octave \fb with $N_\text{oct}=30$ subbands. Thereby, the fullband \ac{RT} estimate $\rteoct$ is determined by the weighted average of the subband estimates $\rtfdeoct$ for $\mu=n_\text{oct},\ldots, N_\text{oct}$ with $n_\text{oct}=20$. The octave \fb design employed for this equals that considered in the \ac{ACE} challenge for the evaluation of the subband \ac{RT}, \cf \cite{Eaton2015}.

Results obtained with the DCT \fb as well as the octave \fb were submitted to the ACE challenge (see \secref{sec:results}).

\presec
\section{Frequency-dependent RT Estimation}
\label{sec:subband_RTE}
\postsec

The estimation of the subband \ac{RT} with an octave \fb is especially difficult for the lower subbands, even if the \ac{RIR} is given (see, \eg \cite{Loellmann11a} and the references cited within). This problem becomes even more pronounced \wrt a blind estimation, and calculating the subband \ac{RT} by means of the baseline algorithm has led to a high estimation error especially for the lower subbands ($1,\ldots,n_\text{oct}-1$). To alleviate this problem, the following approach has been developed: The subband estimates for the lower subbands are extrapolated from the more reliable estimates of the upper subbands  ($n_\text{oct},\ldots,N_\text{oct}$) by means of a model function for the subband \ac{RT}. For this purpose, the following simple model for the frequency-dependent \ac{RT} is devised. Inspection of the ground-truth data for the subband \ac{RT} $\rtfd$ of the Dev database has shown that the frequency-dependency of the subband \ac{RT} can be often roughly approximated by a function similar to that 
of a scaled Rayleigh distribution with an offset $m_0$. This led to the following model function
\begin{align}
 f_\text{mod}(\mu,b) & = \frac{\mu}{\alpha\mul b^2} \exp\left\{\frac{\displaystyle -\mu^2}{\displaystyle 2\mul \alpha^2\mul b^2}\right\} + m_0
\\ 
m_0 & = \frac{1}{N_\text{oct}-n_\text{oct}+1} \sum\limits_{\mu=n_\text{oct}}^{N_\text{oct}} \rtfdeoct
\end{align}
with subband index $\mu$ and $\alpha = 7.5$. The optimal scaling factor is calculated by minimizing the least-square error between the model function and the \ac{RT} estimates for the upper subbands
\begin{align}
b_\text{opt} & = 
\text{arg}\hspace{0.5ex} \minimum{b}{ \sum\limits_{\mu=n_\text{oct}}^{N_\text{oct}} \left|f_\text{mod}(\mu,b)-\rtfdeoct \right|^2 } 
\end{align}
which is calculated here for $n_\text{oct}=20$ and the discrete values $b\in\{0.5,1,1.5,\ldots,5\}$. The subband \ac{RT} estimates are finally given by
\begin{align}
\rtfdemod = f_\text{mod}(\mu,b_\text{opt})
\myforall \mu \in\{\, 1,\ldots,N_\text{oct} \,\}
\eqdot
\end{align}
The developed algorithm is exemplified in \fref{fig:concept_subbandd_RTE}, which shows the \acp{RT} obtained for the speech file \qmarks{\mbox{Single\_Red\_B\_Phil\_Live\_Ambient\_0dB.wav}} of the Dev database.
\begin{figure}[t!]
\newcommand{\scal}{0.9}
\newcommand{\hshift}{-2.8ex}
\psfrag{True RT}[Bl][Bl][\scal]{\sparaise{0.9ex}{0.2ex}{ground-truth\hspace{0.8ex} $\rtfd$}}
\psfrag{Model}[Bl][Bl][\scal]{\sparaise{0.9ex}{0.1ex}{model-based RTE\hspace{0.8ex} $\rtfdemod$}}
\psfrag{Estimated Subband T60}[Bl][Bl][\scal]{\sparaise{0.9ex}{0ex}{direct approach\hspace{0.8ex} $\rtfdeoct$}}
\psfrag{subbands}[Bl][Bl][\scal]{\sparaise{0ex}{0ex}{subband $\mu$}}
\psfrag{T60}[Bl][Bl][\scal]{\sparaise{0ex}{0ex}{RT/\secs}}
\psfrag{0}[Bl][Bl][\scal]{\sparaise{0ex}{0ex}{0}}
\psfrag{0.2}[Bl][Bl][\scal]{\sparaise{0ex}{0ex}{0.2}}
\psfrag{0.4}[Bl][Bl][\scal]{\sparaise{0ex}{0ex}{0.4}}
\psfrag{0.6}[Bl][Bl][\scal]{\sparaise{0ex}{0ex}{0.6}}
\psfrag{0.8}[Bl][Bl][\scal]{\sparaise{0ex}{0ex}{0.8}}
\psfrag{1}[Bl][Bl][\scal]{\sparaise{0ex}{0ex}{1}}
\psfrag{1.2}[Bl][Bl][\scal]{\sparaise{0ex}{0ex}{1.2}}
\psfrag{1.4}[Bl][Bl][\scal]{\sparaise{0ex}{0ex}{1.4}}
\psfrag{1.6}[Bl][Bl][\scal]{\sparaise{0ex}{0ex}{1.6}}
\psfrag{1.8}[Bl][Bl][\scal]{\sparaise{0ex}{0ex}{1.8}}
\psfrag{5}[Bl][Bl][\scal]{\sparaise{0ex}{0ex}{5}}
\psfrag{10}[Bl][Bl][\scal]{\sparaise{0ex}{0ex}{10}}
\psfrag{15}[Bl][Bl][\scal]{\sparaise{0ex}{0ex}{15}}
\psfrag{20}[Bl][Bl][\scal]{\sparaise{0ex}{0ex}{20}}
\psfrag{25}[Bl][Bl][\scal]{\sparaise{0ex}{0ex}{25}}
\psfrag{30}[Bl][Bl][\scal]{\sparaise{0ex}{0ex}{30}}
\sparaise{-2.5ex}{0ex}{\includegraphics[width=1.12\columnwidth ,draft=false ]{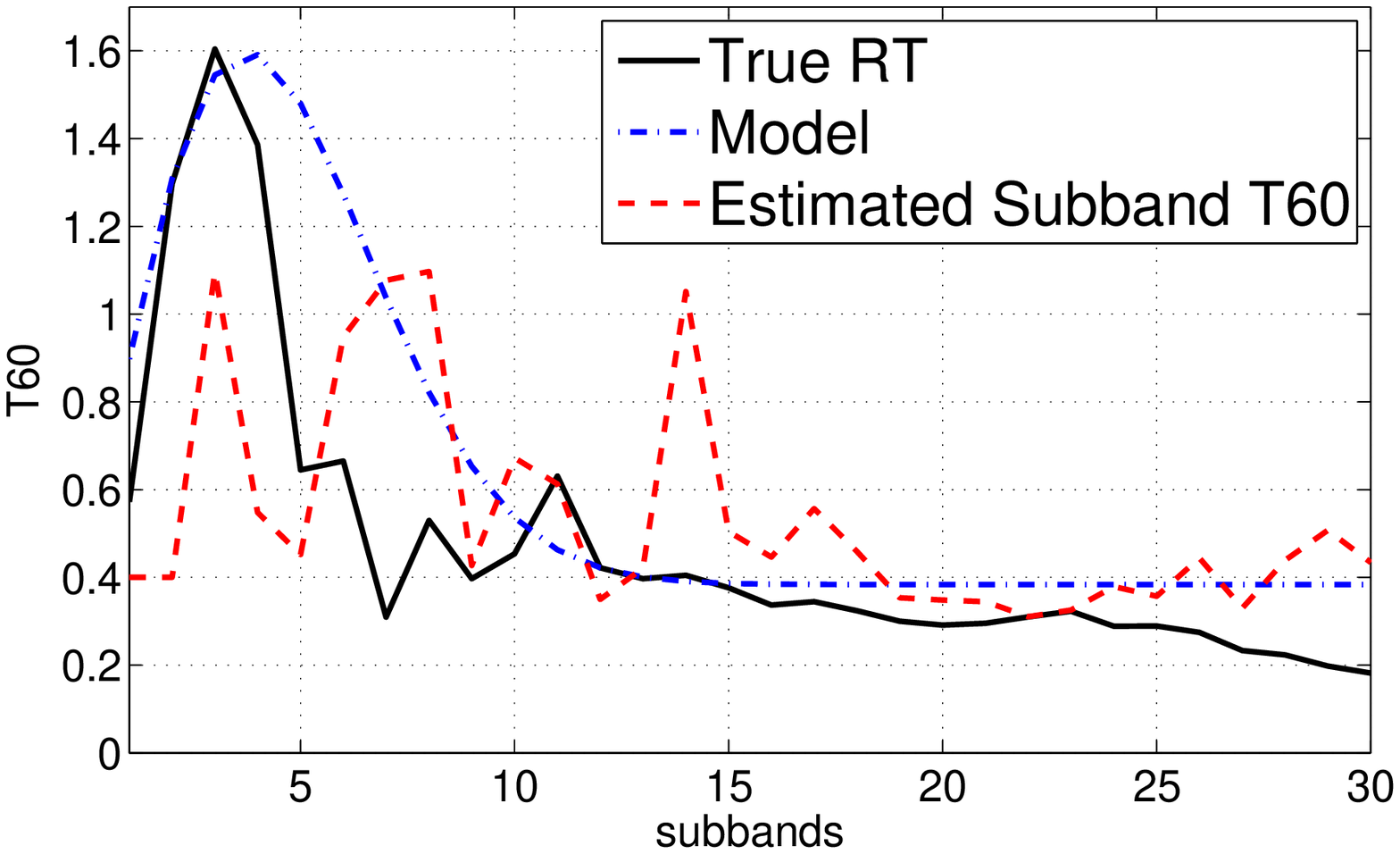}}
\vspace{-3.5ex}
\labeling{fig:concept_subbandd_RTE}{\ac{RT} estimates for 1/3-octave subbands obtained by the new model-based approach and the direct approach where the baseline algorithm is applied for each subband signal.}
\end{figure}
The mean-square error over all subband estimates for this example equals 0.1218 for the direct approach and 0.071 for the model-based \ac{RTE}. Evaluation for the Dev database has on average shown a superior performance for the model-based approach in comparison to the direct approach so that only results for the model-based approach were submitted. It is important to notice that the model-based subband \ac{RTE} also possesses a significantly lower computational complexity than the direct approach.

\presec
\section{Evaluation Results}
\label{sec:results}
\postsec

The evaluation figures provided by \fott{the committee of }the \ac{ACE} challenge are shown in \mbox{\frefs{fig:results_ACE_fan}-\ref{fig:results_ACE_ambient}}. 
\begin{figure}[t!]
\centering
\sparaise{0ex}{0ex}{\includegraphics[width=\fwidth, draft=\draftplot ]{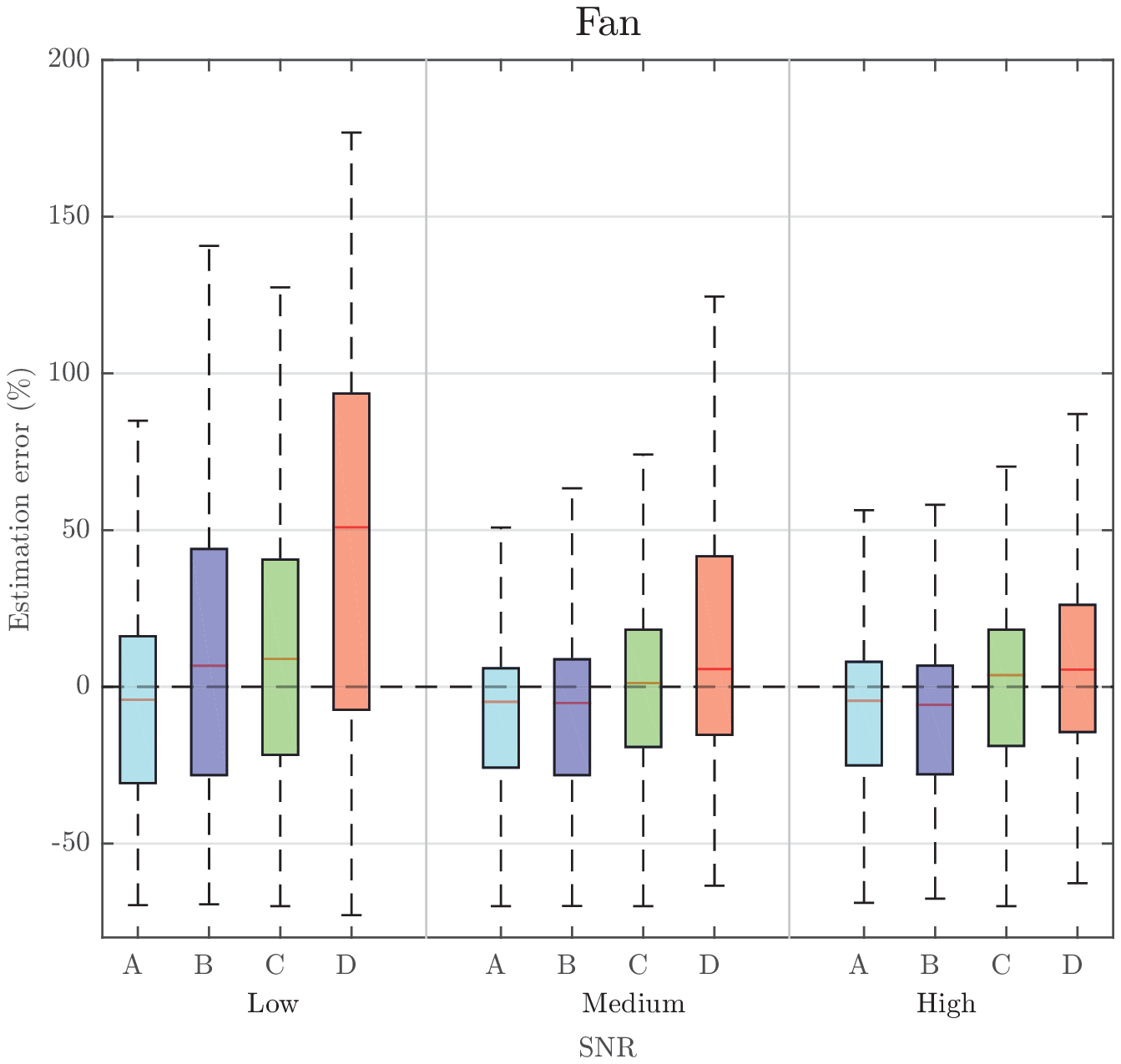}}
\vspalab
\labeling{fig:results_ACE_fan}{Evaluation results of the \ac{ACE} challenge for fan noise: \newline
\hspace*{2ex}\textcolor{ccyan}{\textbf{A}}: \hspace*{1.0ex} \ac{DCT}-based fullband \ac{RTE} $\rtedct$ (1.) \newline
\hspace*{2ex}\textcolor{cpurple}{\textbf{B}}: \hspace*{1.0ex} octave subband-based fullband \ac{RTE} $\rteoct$ (2.) \newline
\hspace*{2ex}\textcolor{cgreen}{\textbf{C}}: \hspace*{1.0ex} baseline algorithm for fullband \ac{RTE} $\rtebase$ (3.) \newline
\hspace*{2ex}\textcolor{cred}{\textbf{D}}: \hspace*{1.0ex} model-based subband \ac{RTE} $\rtfdemod$.} 
\end{figure}
\begin{figure}[t!]
\centering
\sparaise{0ex}{0ex}{\includegraphics[width=\fwidth ,draft=\draftplot ]{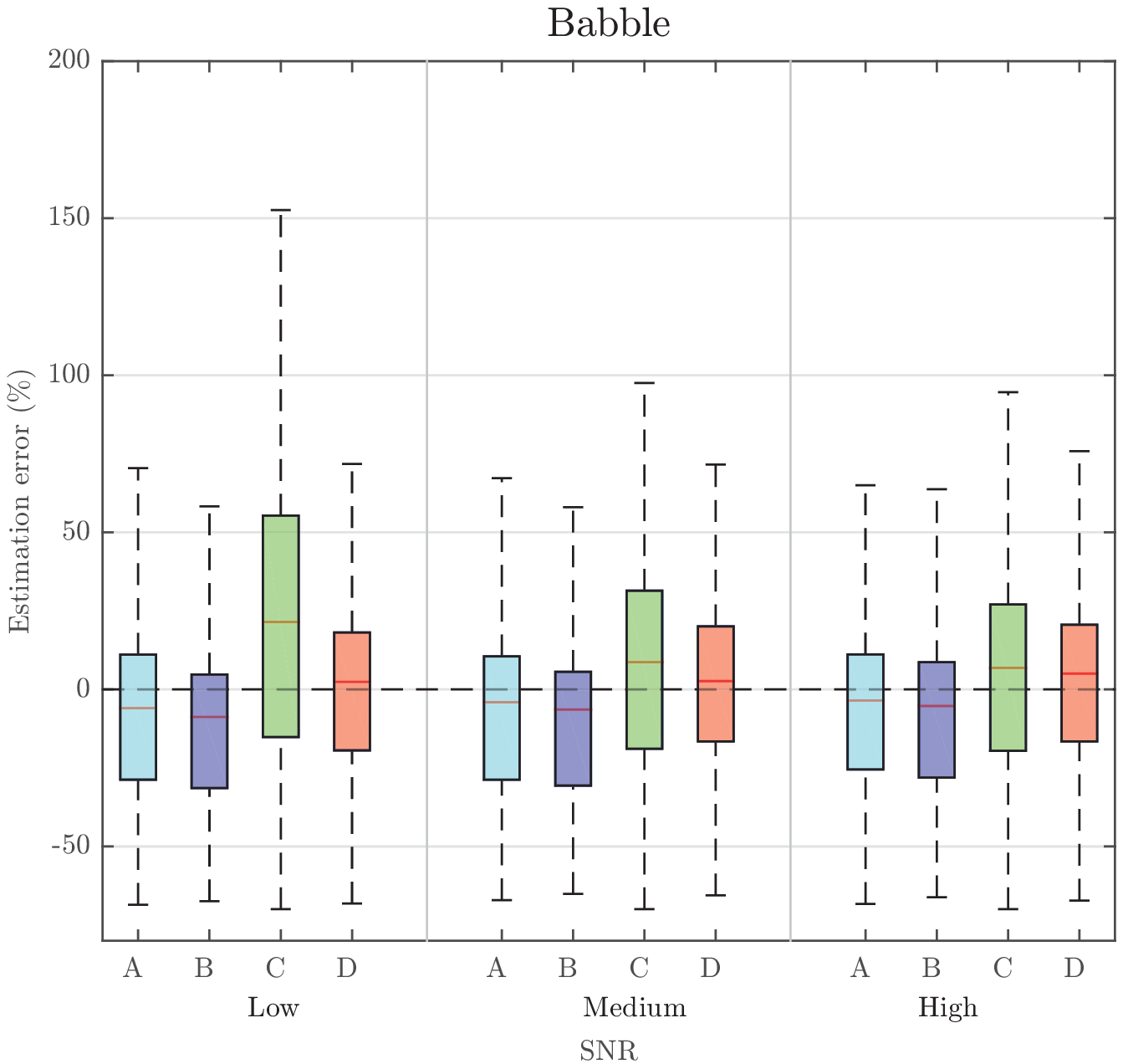}}
\vspalab
\labeling{fig:results_ACE_babble}{Evaluation results of the \ac{ACE} challenge for babble noise (see also \fref{fig:results_ACE_fan}).} 
\end{figure}
\begin{figure}[t!]
\centering
\sparaise{0ex}{0ex}{\includegraphics[width=\fwidth,draft=\draftplot ]{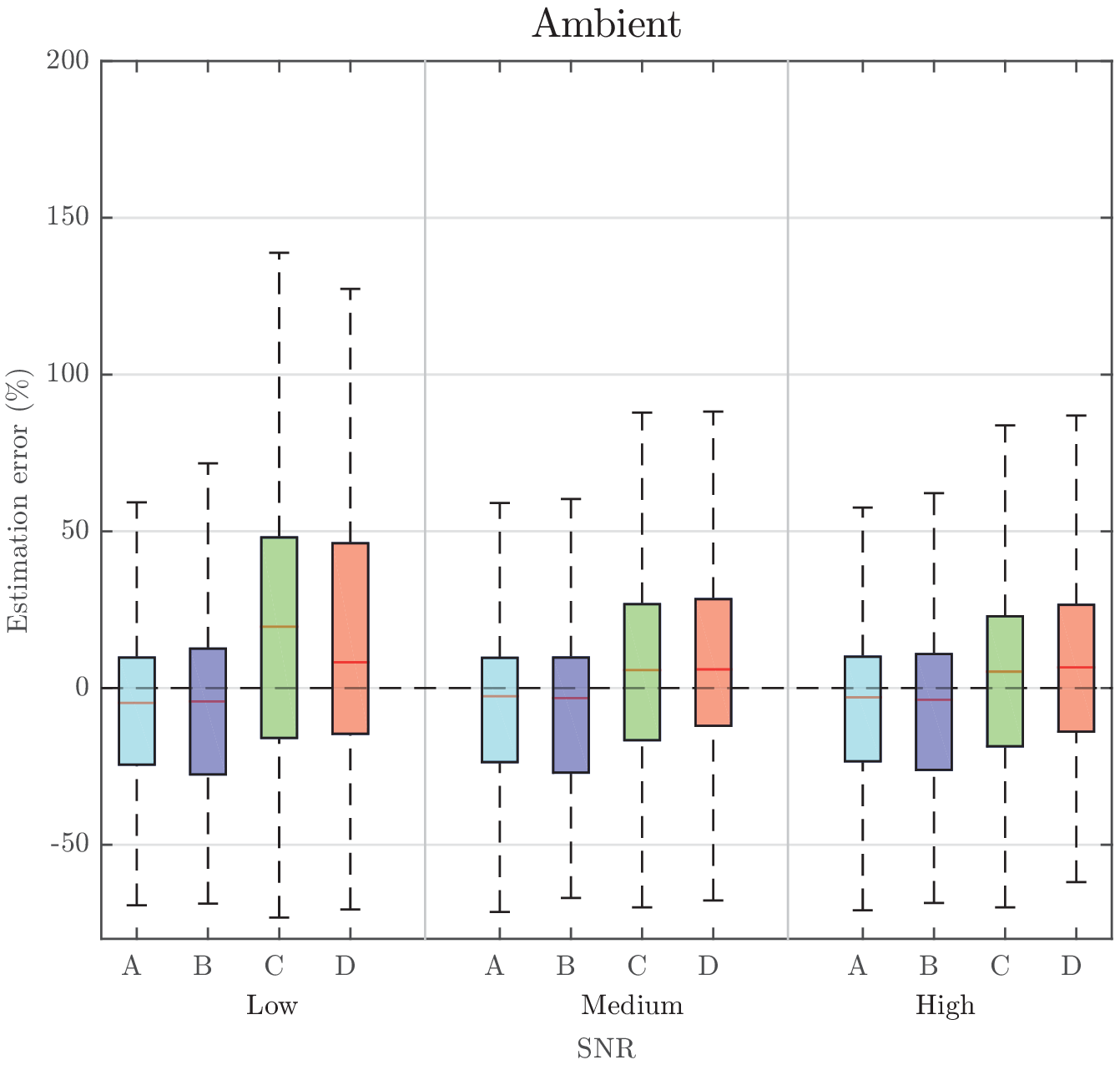}}
\vspalab
\labeling{fig:results_ACE_ambient}{Evaluation results of the \ac{ACE} challenge for ambient noise (see also \fref{fig:results_ACE_fan}).} 
\end{figure}
They display the minimum, first quartile, median, third quartile, and maximum of the relative estimation error $(\rte-\rt)/\rt$ in percent for different noise types and \acp{SNR} of $-1\db$, $12\db$ and $18\db$ for the 4500 speech files of the Eval database. It can be observed that averaging the subband \ac{RT} estimates generally leads to a lower variance for the estimation error and a different bias for the mean value than the baseline approach. Thereby, the \ac{DCT}-based averaging shows mostly a slightly better performance than averaging over octave subband estimates. The superior performance of the \ac{DCT}-based approach over the other two fullband estimators becomes most pronounced for low \acp{SNR}. The priorities assigned to the three algorithms for fullband \ac{RTE} in the submission phase are given in brackets in \fref{fig:results_ACE_fan} and show that the evaluation results comply with the expected performance ranking.  

The model-based subband \ac{RT} estimator achieves, with the exception for low and medium fan noise, a similar performance for the estimation error than the fullband \ac{RT} estimators, even though the blind estimation of the subband \ac{RT} can be considered to be more challenging than the estimation of the fullband \ac{RT}.

\presec
\section{Conclusions}
\label{sec:conclusions}
\postsec

The presented baseline algorithm allows to estimate the fullband \ac{RT} with low complexity (\cf \cite{Loellmann2010b}). The proposed weighted averaging of the subband \ac{RT} estimates\fott{ in the \ac{DCT} or octave subband domain} reduces the estimation variance at the price of an increased computational complexity. The devised model-based subband \ac{RTE} achieves an estimation error, which is for most scenarios in a similar range as for the presented fullband algorithms. 

For all treated algorithms, the estimation errors are mainly caused by the fact that the underlying statistical model for the \ac{RIR} with an exponential decay according to \eref{eq:RIR_model} is often not valid, especially at high \acp{DRR}. Therefore, the search for blind estimation algorithms with an improved, more appropriate model for the reverberant sound decay remains a promising direction for further work.